\documentclass[journal=langd5,manuscript=article]{achemso}

\usepackage{graphicx}

\newcommand{\eq} {{\rm eq}}

\newcommand{\rmb} {{\rm b}}
\newcommand{\rmB} {{\rm B}}

\newcommand{\taud} {\tau_{\rm d}}
\newcommand{\tauk} {\tau_{\rm k}}

\title{Dynamic Surface Tension of Aqueous Solutions of Ionic Surfactants: Role of Electrostatics}

\author{Hern\'an Ritacco} \affiliation{Departamento de Qu\'imica F\'isica I, Facultad de Cs. Qu\'imicas,
    Universidad Complutense de Madrid, 28040 Madrid, Spain}

\author{Dominique Langevin}
\email{langevin@lps.u-psud.fr}
\affiliation{Laboratoire de Physique des Solides, Universit\'e Paris Sud, 91405 Orsay cedex, France}

\author{Haim Diamant}
\affiliation{Raymond \& Beverly Sackler School of Chemistry, Tel Aviv University, Tel Aviv 69978, Israel}

\author{David Andelman}
\affiliation{Raymond \& Beverly Sackler School of Physics \& Astronomy, Tel Aviv University, Tel Aviv 69978, Israel}
\date{Nov 5, 2010}
\begin{document}

\maketitle
\baselineskip=14pt
\begin{abstract}
The adsorption kinetics of the cationic surfactant dodecyltrimethylammonium bromide at the air-water interface has been
studied by the maximum bubble pressure method at concentrations below the critical micellar concentration. At short times, the adsorption is diffusion-limited. At longer times, the surface tension shows an intermediate plateau and can no longer be accounted for by a diffusion limited process. Instead, adsorption appears kinetically controlled and slowed down by an adsorption barrier. A Poisson-Boltzmann theory for the electrostatic repulsion from the surface does not fully account for the observed potential barrier. The possibility of a surface phase transition is expected from the fitted isotherms but has not been observed by Brewster angle microscopy.
\end{abstract}

\section{Introduction}

Over the last decades, the adsorption kinetics of surfactants at fluid interfaces has been the subject of many studies due to its prime importance in numerous applications such as wetting, detergency, foaming and emulsification\cite{Miller_book}. Adsorption kinetics is commonly studied by creating a freshly exposed surface in contact with
a bulk surfactant solution and measuring the temporal dependence of its surface tension. This dynamic surface tension\cite{d1,d2} has been shown to correlate much better than the equilibrium surface tension with properties crucial for applications, such as foaming ability of surfactant solutions and the spreading velocity of fluid films on top of solid substrates\cite{d3,d4}.

Experimental work on dynamic surface tension stimulated over the years
theoretical modeling. In their pioneering model from the 1940s, Ward
and Tordai\cite{d5} assumed that the adsorption is limited by
diffusion (DLA), resulting in an asymptotic decay as the inverse
square root of time. The Ward-Tordai model is still widely used
nowadays to analyze dynamic surface tension data. Yet, the improved
accuracy achieved in recent years has revealed important deviations
from the DLA behavior in several systems\cite{d6}. These deviations
have been attributed to a number of factors such as the existence of
kinetic adsorption barriers, especially for charged systems\cite{d1};
the role of surface-active impurities\cite{d7}; lateral relaxation
modes in the monolayer plane\cite{d8}; and the formation of surface
crystals\cite{d9}. { We note that the effect of electrostatic
  barriers on the kinetics of surfactant aggregation was analyzed also
  in the different context of ionic micelles \cite{Aniansson76}.}

For example, an anomalous long time decay with an intermediate plateau
has been observed for the ionic solutions of sodium dodecylsulfate
(SDS) with no added salt. Since surfactant hydrolysis produces trace
amounts of dodecanol that slowly adsorb at the water-air interface,
some authors attributed the peculiar surface tension kinetics to these
surface-active impurities\cite{d7}. A similar anomalous behavior was
reported for aqueous solutions of aerosol OT\cite{d3,d10}, as well as
for SDS at the alkane-water interface.\cite{d11} In the latter case,
dodecanol impurities are not expected to accumulate at the interface
and to affect the adsorption dynamics because they are soluble in the
alkane phase\cite{d11}.

While the origin of non-DLA-like behavior is not well understood for
all systems, there is strong evidence that the adsorption of pure
nonionic surfactants is usually
diffusion-limited\cite{HD_jpc96,HD_csa01}. For ionic surfactant
solutions, however, an electrostatic surface potential is
progressively created as the adsorption proceeds. This surface
potential acts as an adsorption barrier for additional surfactant
molecules as they migrate from the bulk toward the surface, thus
potentially giving rise to a non-DLA behavior.  The dynamic surface
tension of ionic surfactants (SDS) at the oil-water interface was
previously shown to be different from DLA behavior\cite{anne}, but
when salt was added to the ionic surfactant solution, the process
returned to a DLA-like behavior. This can be explained by
electrostatic screening of the surface potential in the presence of
added electrolyte.

In the present work we extend the results of Ref~\citenum{anne} to the
adsorption of ionic surfactants at the air-water interface. We have
studied aqueous solutions of the cationic surfactant
dodecyltrimethylammonium bromide (DTAB) at various concentrations
below the critical micellar concentration ({\em cmc}) using the
maximal bubble pressure (MBP) method. At short times the kinetic
behavior is found to be DLA-like, while at longer times the kinetics
appears to be slowed down by an adsorption barrier. The experimental
results are analyzed and compared with the theoretical predictions of
Ref~\citenum{HD_jpc96}. The influence of the adsorption barrier is better seen just before the critical micellar concentration ({\em cmc}) and working below the {\em
  cmc} has the advantage that the theoretical interpretation is more
straightforward as it does not have to take into account the presence
of micelles. In the SDS case, hydrolysis produces dodecanol, which is very surface active at these concentrations (at oil-water interfaces it is less a problem, because dodecanol is solubilised in the oil phase). We have chosen here to work with DTAB instead of SDS, because DTAB is much more stable chemically.

\section{Materials and Methods}
Dodecyltrimethylammonium bromide (DTAB) was supplied by Sigma ($>99\%$
purity) and used without further purification. De-ionized water was
obtained from a DEIONEX MS 160 equipment, with a resistivity of
12\,M$\Omega$. All the measurements were performed at $25 \pm 0.2
^{\circ}$C, and the solutions were prepared by dilution of
concentrated stock solutions.  The equilibrium surface tension
measurements were conducted using a circular Teflon trough (capacity
5\,ml), housed in a Plexiglas box with an open-frame version of the
Wilhelmy plate. We used a rectangular open frame ($20\times 10$\,mm),
made of a platinum wire and attached to a force transducer mounted on
a motor, allowing it to be drawn away from the surface at a controlled
constant rate\cite{Atef}.

The dynamic surface tension was measured by a maximum bubble pressure
(MBP) method. This in-house-made instrument (described in
Ref~\citenum{MBP}) is used to measure the maximum pressure necessary
to detach a bubble from a capillary. The surface tension $\gamma$ is
then obtained from the Young-Laplace equation, $\gamma = {R \Delta
  P}/{2}$, where $\Delta P$ is the maximum pressure difference between
the gas in the bubble and the surrounding liquid, and $R$ is the
internal radius of the capillary. We have used a glass capillary with
an inner diameter of 200 $\mu$m, hydrophobized with
hexamethyldisilazane (Sigma). The advantage of such a device over
methods based on analysis of the bubble (or drop) shape, is that it
allows access to short adsorption times, down to about 10\,ms, whereas
the shape methods are limited to times longer than 1 or 2\,s.  The MBP
instrument was tested with the non-ionic surfactant Triton X100, and
the results were found to be compatible with DLA kinetics, indicating
that uncontrolled convection effects were negligible even at times
down to $0.1$\,s.

With the MBP instrument, the bubble surface age is usually associated
with the time interval between consecutive detachments of two bubbles.
However, there is a dead time (delay) between the detachment of a
bubble and the formation of the next bubble. In a recent paper,
Christov et al.\cite{nikolay} showed that the effective aging time of
the interface, $t$, is smaller than and proportional to the time
actually measured between consecutive detachments, $t_{\rm age}$: $t =
t_{\rm age} \lambda^{-2}$, where $\lambda$ is an apparatus constant
described in more details in Ref~\citenum{MBP}.  From the experiments
with Triton X-100 and its diffusion constant ($D=4\times
10^{-6}$\,cm$^2$/s) the apparatus constant, $\lambda = 4.3 \pm 0.3$,
is obtained.  In the following, all the curves are represented as a
function of the corrected time, $t$.

Since it is known that different MBP experimental setups give different dynamic surface tension
curves $\gamma(t)$\cite{fainerman}, experiments were also performed with a commercial MBP instrument (MPT2 from Lauda) for comparison with the in-house made apparatus.
We also employed a Brewster angle microscope (Mini-BAM from NFT) together with a Langmuir trough (NIMA 601BAM) to image the surface monolayer and visualize any of its possible surface heterogeneities.

\section{Results}
\begin{figure}[tbh]
\vspace{.8cm} \centerline{\resizebox{0.6\textwidth}{!}
{\includegraphics{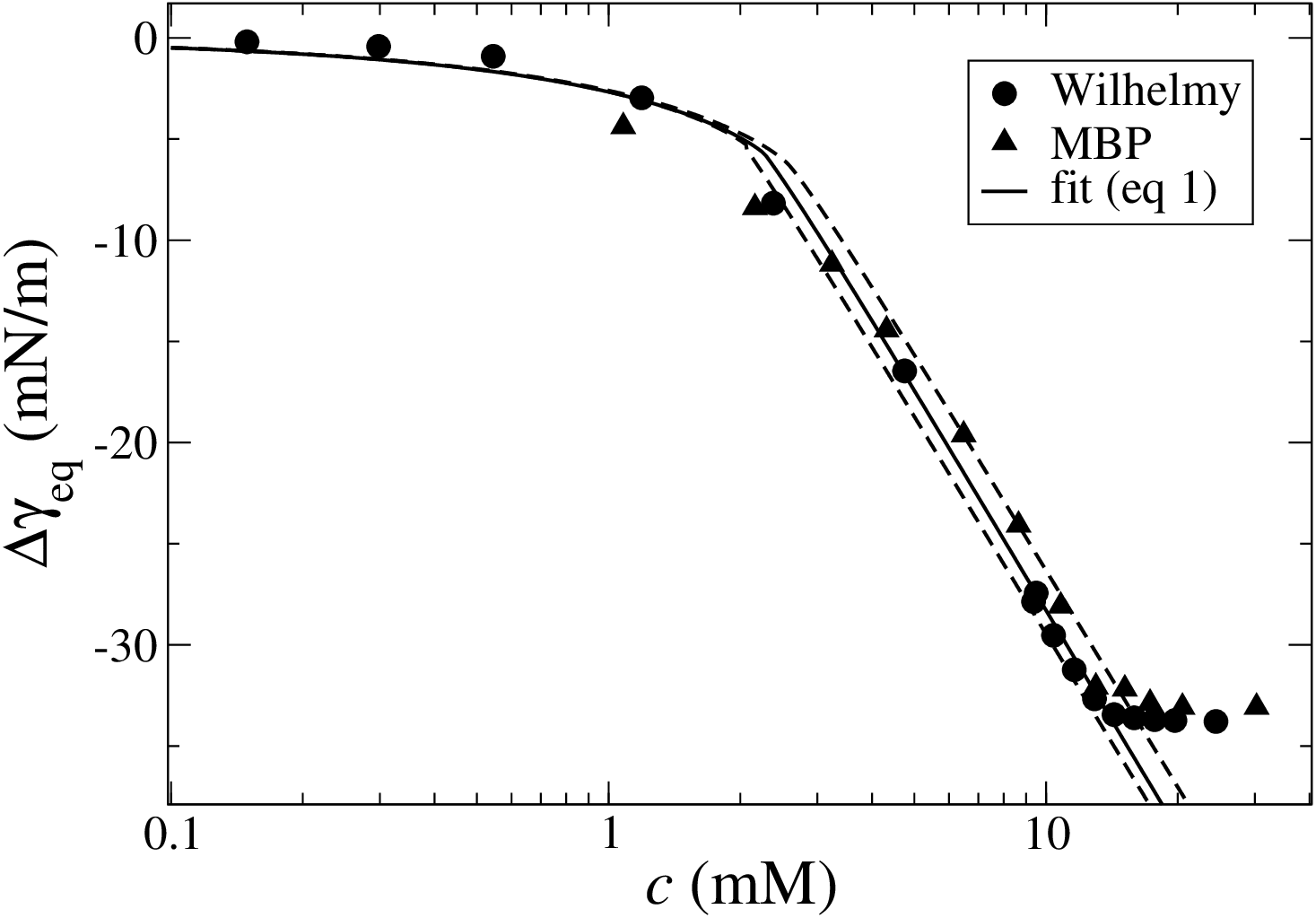}}} \caption[]{\textsf{Equilibrium surface
tensions obtained by the Wilhelmy plate (circles) and MBP
(up-triangles) techniques. The solid line is a fit obtained by using
eq~\plainref{gamma_eq}. The values of the fit parameters are:
$\alpha=11.8$, $\beta=7.6$, and $a=7.2$\,\AA. { The upper and lower
dashed lines show the theoretical curves using the same values for
$\alpha$ and $a$ but with $\beta=7.2$ and $8.0$, respectively}.}}
\label{tens_eq}
\end{figure}

\ref{tens_eq} shows the reduction in the equilibrium surface tension,
$\Delta\gamma_{\rm eq}=\gamma_{\rm eq}-\gamma_{\rm w}$, for salt-free
DTAB solutions as a function of the DTAB concentration $c$, where
$\gamma_{\rm w}$ is the bare air-water surface tension. The data
obtained by the two methods --- the Wilhelmy plate and the MBP at long
times (all measurements reach the equilibrium values for $t_{\rm age}
< 50$\,s) --- are in good agreement.  For concentrations above the
{\em cmc} ($ =15$\,mM), the surface tension saturates to a constant
value, $\Delta\gamma_{\rm eq} \simeq -33$\,mN/m.

\begin{figure}[t!]
\vspace{.8cm} \centerline{\resizebox{0.4\textwidth}{!}
{\includegraphics{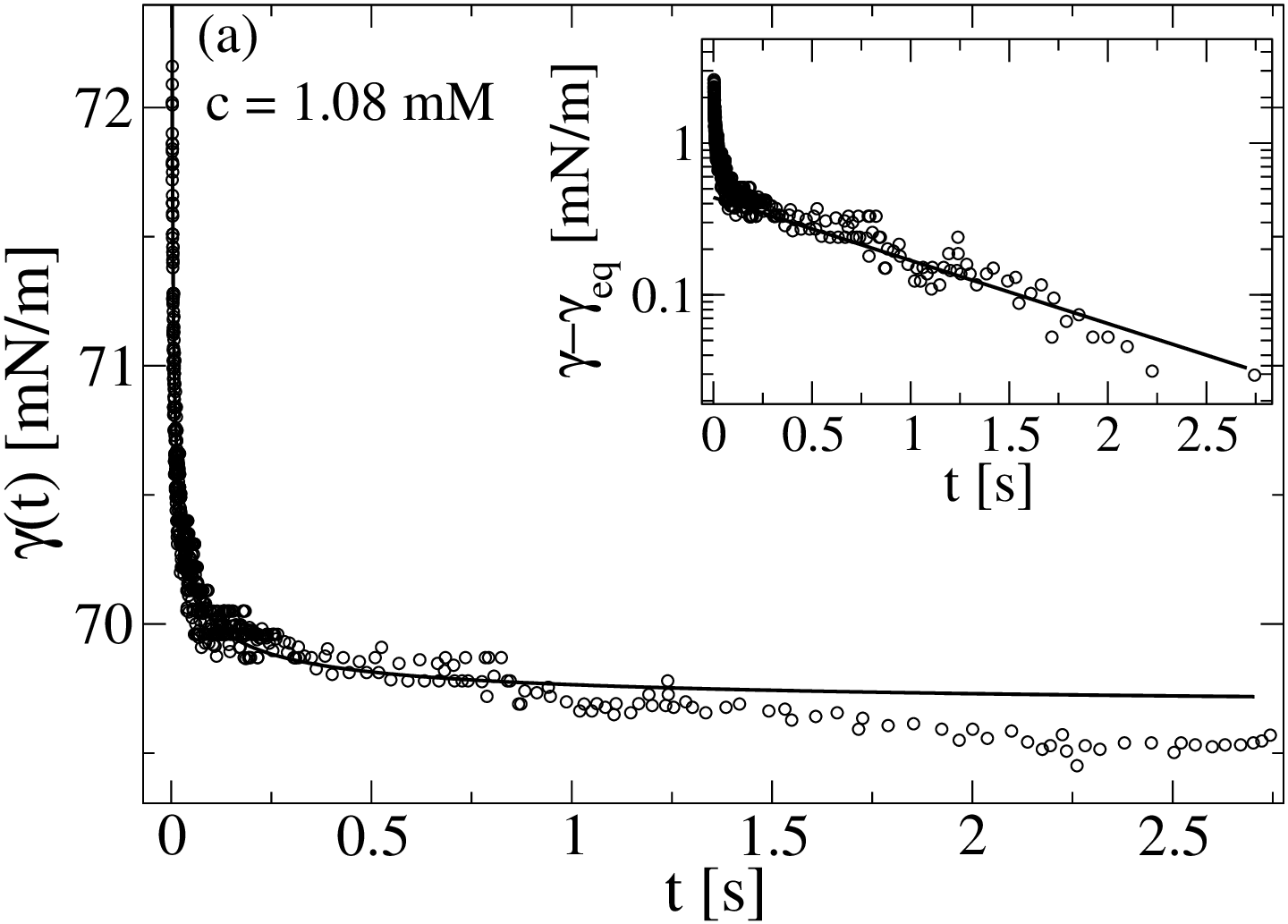}} \hspace{0.3cm}
\resizebox{0.4\textwidth}{!} {\includegraphics{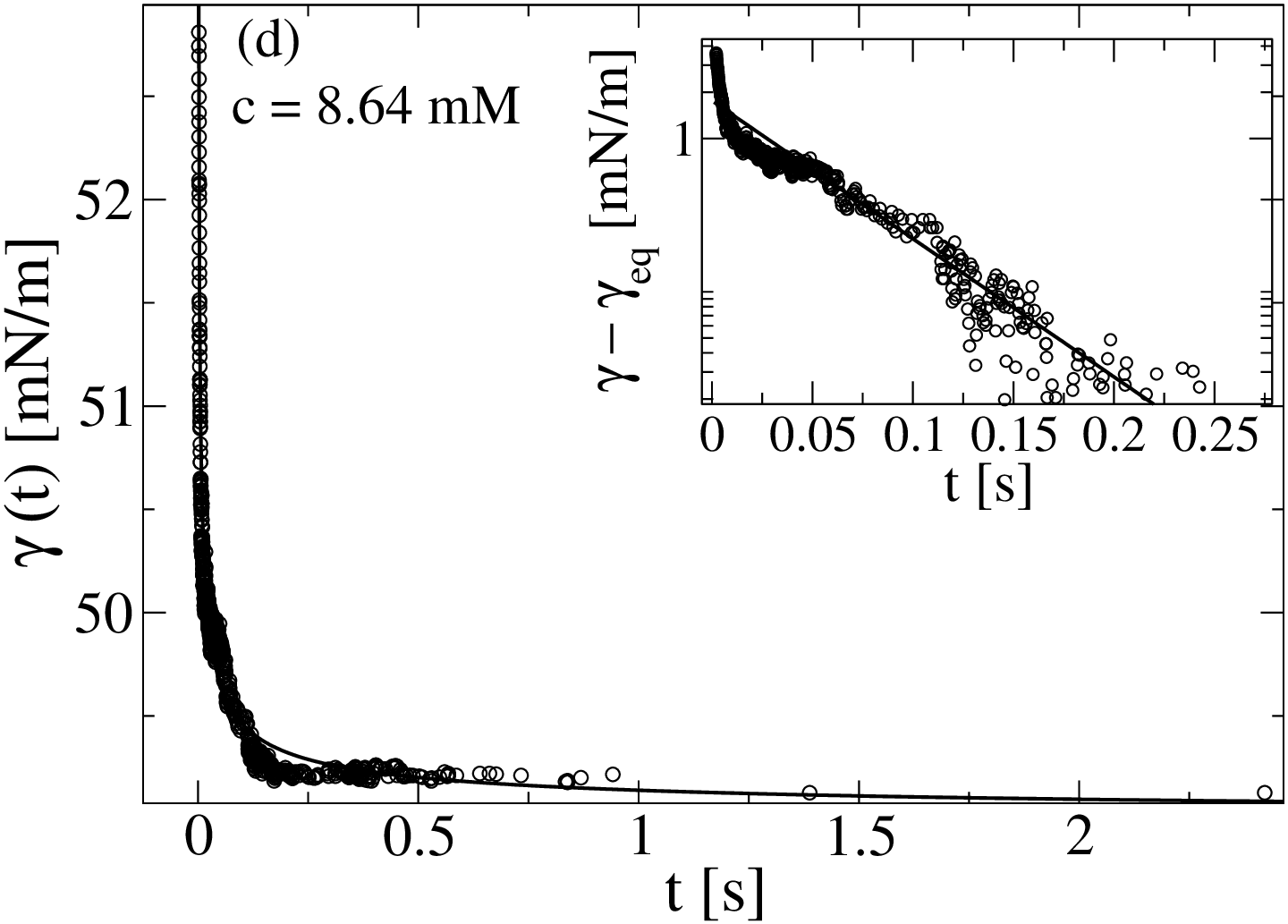}}}
\vspace{0.8cm} \centerline{\resizebox{0.4\textwidth}{!}
{\includegraphics{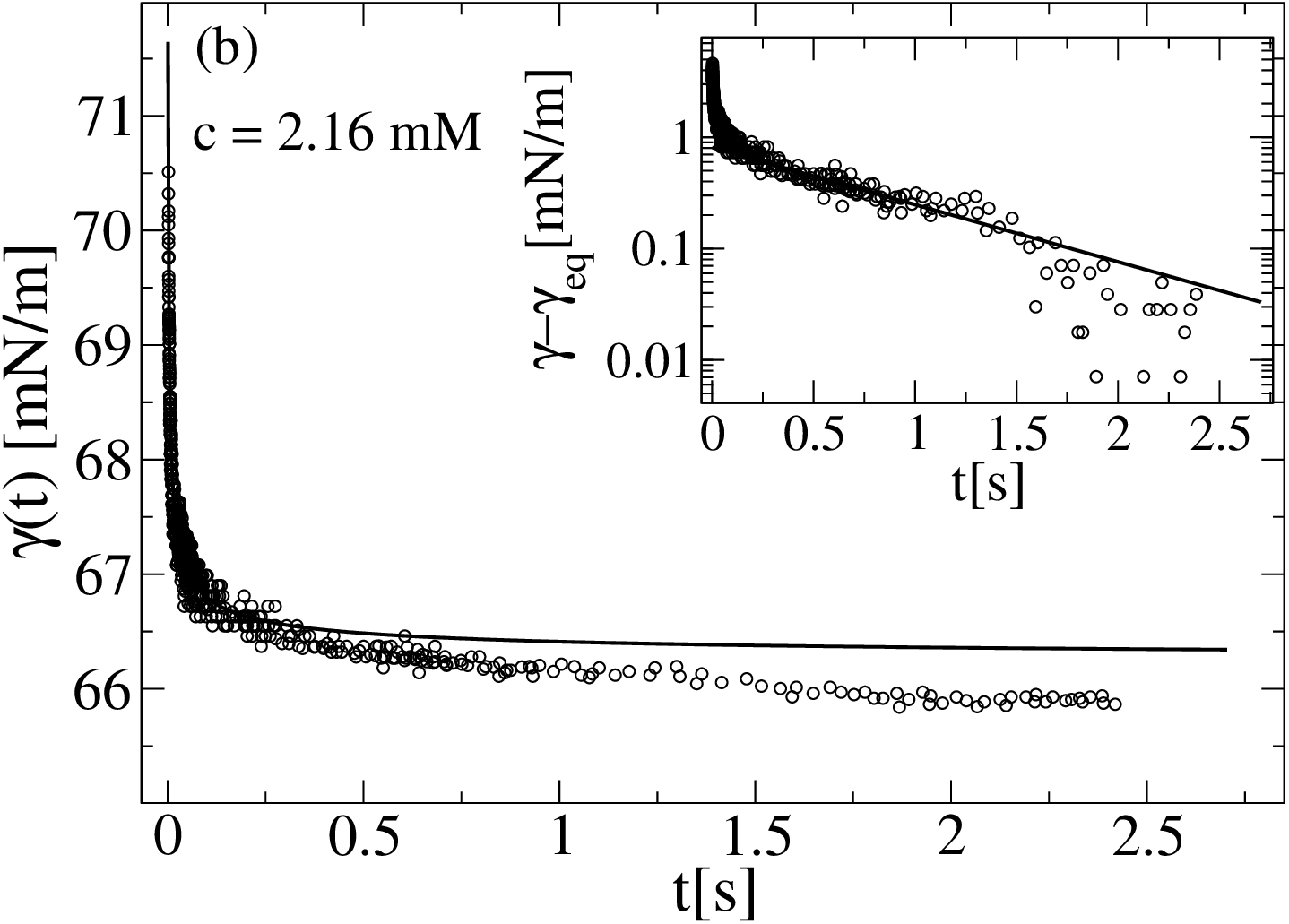}} \hspace{0.3cm}
\resizebox{0.4\textwidth}{!} {\includegraphics{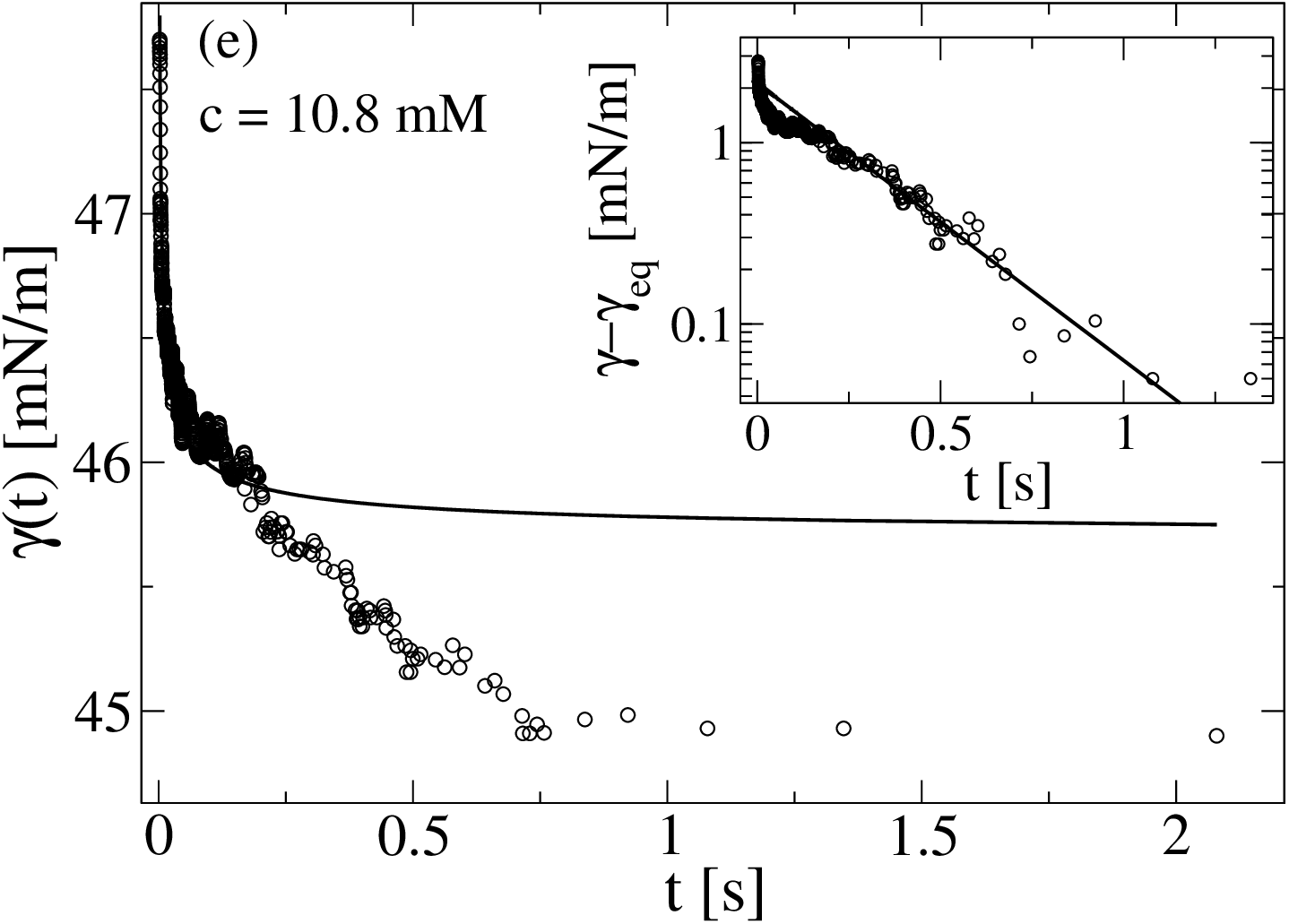}}}
\vspace{0.8cm} \centerline{\resizebox{0.4\textwidth}{!}
{\includegraphics{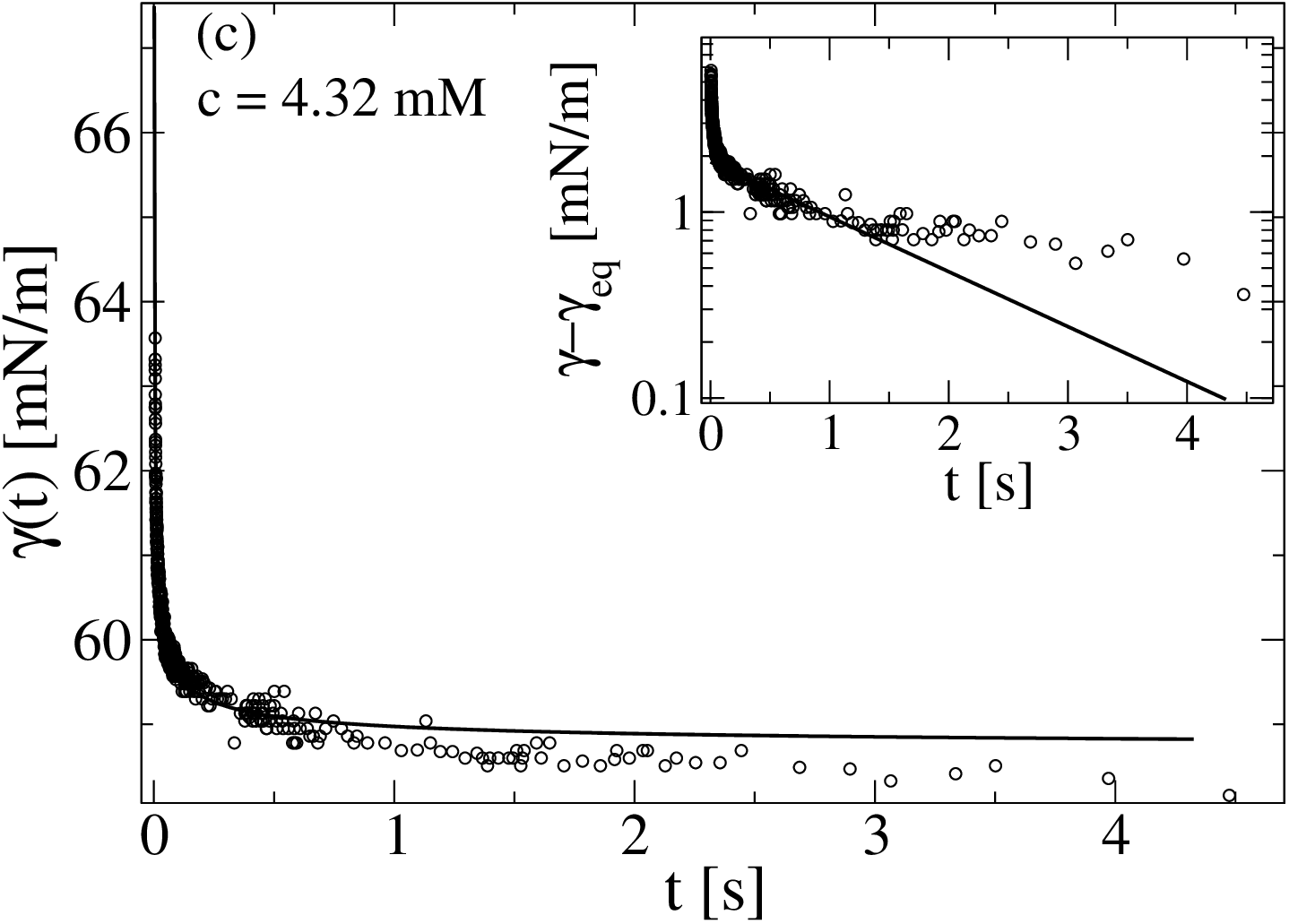}} \hspace{0.3cm}
\resizebox{0.4\textwidth}{!} {\includegraphics{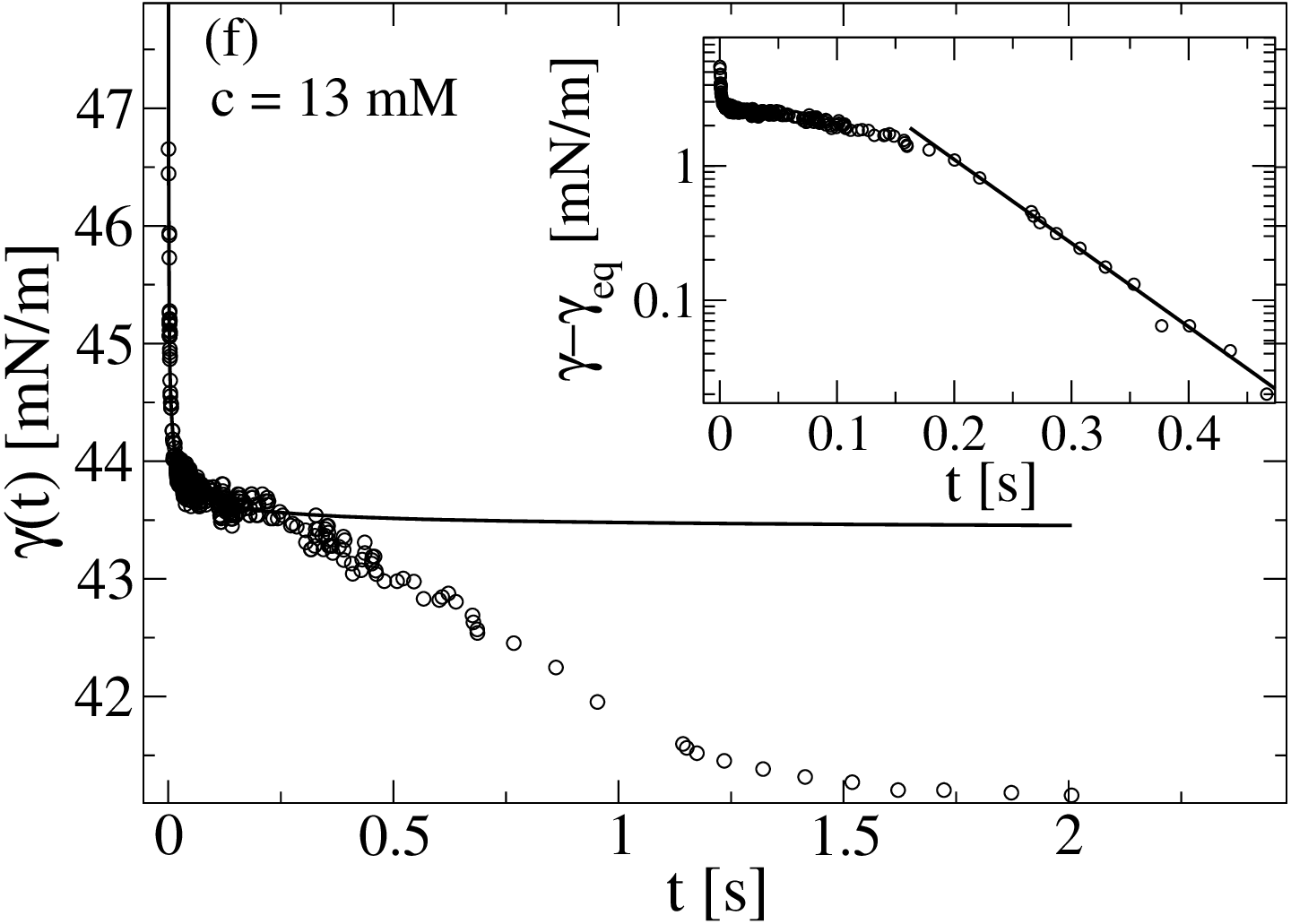}}}
\caption{\textsf{Dynamic surface tension for DTAB concentrations
of (a) $c=$ 1.08 mM, (b) $c=$ 2.16 mM, (c) $c=$
4.32 mM, (d) $c=$ 8.65 mM, (e) $c=$ 10.8 mM ,
and (f) $c=$ 13 mM. The lines are fits to eq~\plainref{diff}, while
in the insets that data points are fitted with eq~\plainref{kinetic}.
}}
\label{c124}
\end{figure}

\ref{c124} shows dynamic surface tension curves for DTAB
concentrations below the {\em cmc}. { At the higher concentrations
(panels d--f) the curves manifest an intermediate plateau indicating a
double-relaxation process.} The behavior observed for high DTAB
concentrations at the air-water interface is qualitatively similar to
that found for SDS at oil-water interfaces at lower SDS concentrations
(the {\em cmc} is smaller for SDS)\cite{anne}. In the latter case,
however, the characteristic times were longer (because the
concentrations were smaller) and accessible by the drop shape method.

The plateaus seen could correspond to the occurrence of a phase
transition in the surface layer during which surface pressure $\Pi =
\gamma - \gamma_w $ remains constant\cite{trans2D}.  In order to check
for this possibility, we imaged, using a BAM apparatus, the surface at
surfactant concentrations close (below and above) the ``knee'' ($c
\simeq 2$\,mM) in~\ref{tens_eq}, but no monolayer heterogeneities were
observed (images not included). Note, however, that this does not
exclude the presence of domains with sizes below the optical
resolution ($\approx 5 \times 10^{-7}$\,m).

\begin{figure}[tbh]
\vspace{0.7cm} \centerline{\resizebox{0.6\textwidth}{!}
{\includegraphics{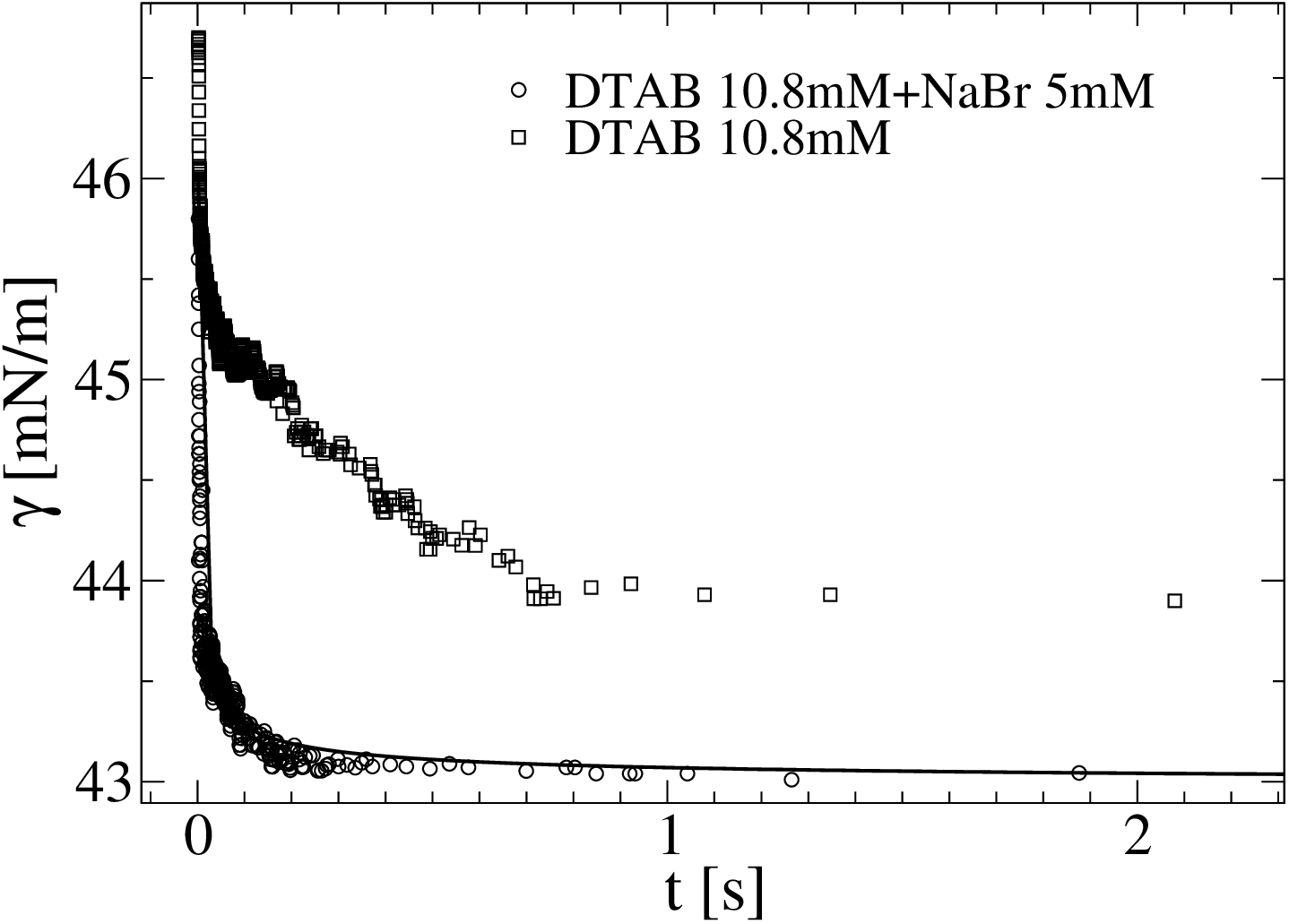}}}
\caption[]{\textsf{Dynamic surface tension for a mixed solution of 10.8\,mM DTAB and 5\,mM NaBr (open
circles), compared with the corresponding salt-free DTAB solution (open squares).
The solid line is a fitting curve using the DLA model of eq~\plainref{diff}. }}
\label{salt}
\end{figure}

\ref{salt} shows a comparison of the dynamic surface tension of
solutions containing $10.8$\,mM DTAB without salt (same
as~\ref{c124}f) and in the presence of 5\,mM NaBr. Placing the two
curves on the same figure shows that the small amount of added salt
has almost suppressed the plateau. In fact, the line in that figure
represents a fitting to a DLA process using eq~\plainref{diff}. From
this fitting and eq~\plainref{a4D} we obtain a diffusion coefficient
$D\simeq 7\times 10^{-6}$\,cm$^2$/s, which is rather close to the
previously measured one,~\cite{DTABdiff} $D\simeq 6\times
10^{-6}$\,cm$^2$/s. The difference between these two values of $D$ is
compatible with uncertainties in the apparatus constant, $\lambda$.

\begin{figure}[tbh]
\vspace{0.7cm} \centerline{\resizebox{0.6\textwidth}{!}
{\includegraphics{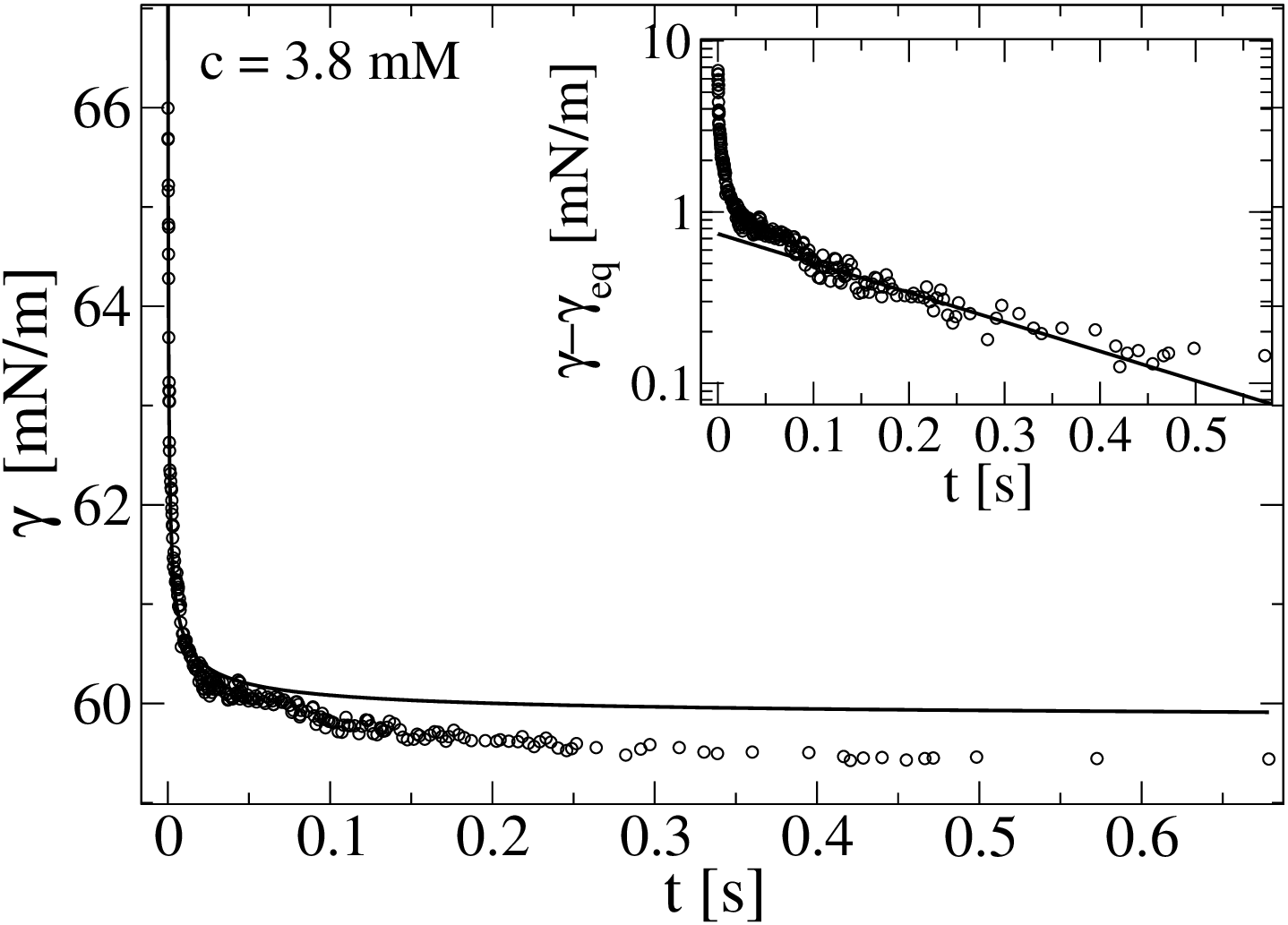}}}
\caption[]{\textsf{Dynamic surface tension for 3.8\,mM DTAB solution measured with a
    commercial MBP apparatus (MPT2). The solid line is a fitting with
    eq~\plainref{diff}. In the inset, the fitting is done with
    eq~\plainref{kinetic}. Note that the equilibration times are about
    10 times shorter than those of \ref{c124}c.}}
\label{MPT2}
\end{figure}

In~\ref{MPT2} we show the dynamic surface tension for $c = 3.8$\,mM
obtained with the commercial MBP apparatus (MPT2). The results are in
qualitative agreement with our in-house apparatus. Note that the
plateau can equally be observed, but the characteristic times are much
shorter (as compared with~\ref{c124}). For this commercial MBP the
apparatus constant $\lambda$ seems to be lower than 1 (probably due to
some uncontrolled convection). Therefore, we did not analyze
quantitatively these results.

\section{Data Analysis}
\subsection{Equilibrium Data}
\label{sec_equil}

The equilibrium adsorption isotherm and equation of state for ionic
surfactant solutions were previously
derived\cite{HD_jpc96,HD_csa01}. They are given by
%
\begin{eqnarray}
  \phi_{0,\eq} &=& \frac {\phi_\rmb} {\phi_\rmb + [b\phi_{0,\eq} +
  \sqrt{(b\phi_{0,\eq})^2 + 1}]^2 {\rm e}^{-\alpha -\beta\phi_{0,\eq}}}
 \nonumber\\
  \Delta\gamma_\eq &=& \frac{k_\rmB T}{a^2} \left[ \ln(1-\phi_{0,\eq})
  + \frac{\beta}{2}
  \phi_{0,\eq}^2 - \frac{2}{b}\left( \sqrt{(b\phi_{0,\eq})^2+1}-1 \right)\right].
\label{gamma_eq}
\end{eqnarray}

In eq~\plainref{gamma_eq}, $\phi_\rmb=a^3 c$ is the bulk surfactant volume
fraction ($a$ being the average size of a surfactant molecule),
$\phi_{0,\eq}$ is the surfactant area fraction (surface coverage) at
equilibrium, and $k_\rmB T$ the thermal energy. The parameters
$\alpha$ and $\beta$ are the Langmuir adsorption parameter and the
Frumkin lateral interaction parameter, respectively (both given in
units of $k_\rmB T$). Finally, the parameter $b(a,\phi_\rmb)=[\pi
l_\rmB/(2a\phi_\rmb)]^{1/2}$, where the Bjerrum length,
$l_\rmB=e^2/(\varepsilon k_\rmB T)\simeq 7$ \AA, characterizes the
strength of electrostatic interactions, and $\varepsilon$ is the water
dielectric constant.

We numerically solve the isotherm of eq~\plainref{gamma_eq} for
$\phi_{0,\eq}$ and substitute the result in the equation of state to
calculate $\Delta\gamma_\eq$. \ref{tens_eq} shows the fit of $\Delta
\gamma_{\rm eq}(c)$, { thus obtained}, to the equilibrium data (up to
the {\em cmc}).  There are three parameters in the fit: $\alpha$,
$\beta$ and $a$. The theoretical curve shown in~\ref{tens_eq}
corresponds to a fitting with $\alpha=11.8 \pm 0.2$, $\beta=7.6 \pm
0.4$, and $a=7.2 \pm 0.2$\,\AA. { In \ref{tens_eq} we also demonstrate
  the sensitivity of the fit to the value of the Frumkin lateral
  interaction parameter, $\beta$.} The key point is that a relatively
large value of $\beta$ is required to reproduce the ``knee'' in
$\Delta\gamma_\eq(c)$ at intermediate concentrations. The need for
such a strong surfactant--surfactant attraction at the interface for
ionic surfactants was recognized before\cite{HD_jpc96}, where it was
suggested that it may be related to the adsorption of a small amount
of counterions, which reduces the electrostatic repulsion, permitting
the interaction of the non-polar parts of the surfactant chains.

\begin{figure}[tbh]
\vspace{0.8cm} \centerline{\resizebox{0.6\textwidth}{!}
{\includegraphics{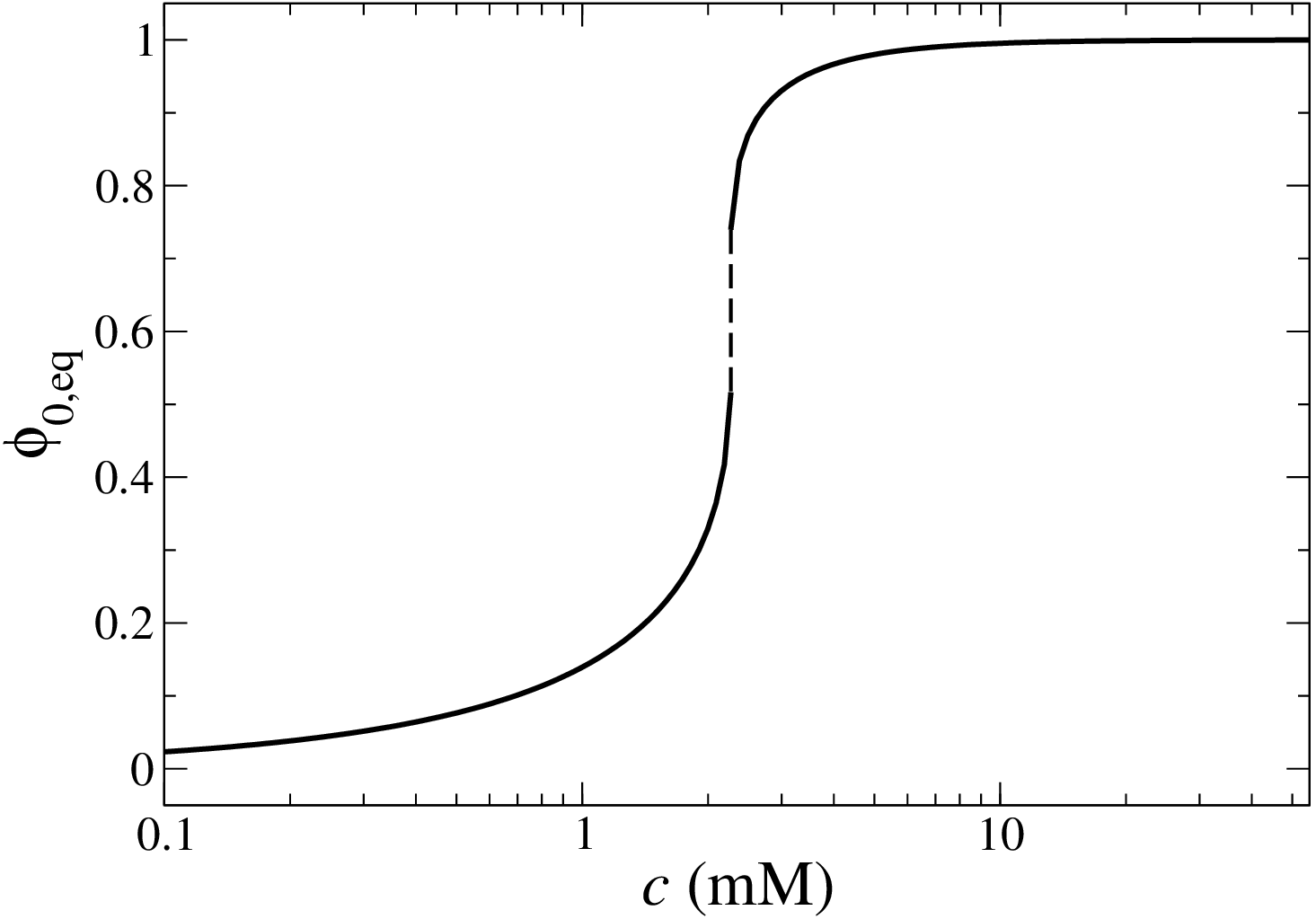}}} \caption[]{\textsf{Theoretically
calculated equilibrium surface coverage, $\phi_{0,\eq}$, as a function of bulk
concentration, $c$. The curve is obtained from eq~\plainref{gamma_eq} with the
parameter values as used to fit~\ref{tens_eq}. The dashed line section
indicates a region of discontinuous (first-order) phase transition for
$0.52\le\phi_{0,\eq}\le0.74$.}} \label{phi_0}
\end{figure}

When $\beta > 7.5$ the fit of the model with the equilibrium data
predicts a surface phase transition above a certain critical surface
coverage. Since the surfactant molecules are soluble in the bulk
solution, the surfactants at the surface can be treated as in a
grand-canonical ensemble, and {\em at equilibrium} there cannot be a
coexistence region between dilute and dense domains.  At a certain
value of the chemical potential (or equivalently of $\phi_\rmb=a^3c$)
$\phi_{0,\eq}$ should jump discontinuously without a change in
$\gamma$.  {\it Kinetically}, however, this increase in the coverage
should proceed via nucleation and growth of domains of the denser
phase.  For the above-mentioned values of parameters $\alpha$, $\beta$
and $a$, the transition is between $\phi_0\simeq 0.52$ (dilute) and
$\phi_0\simeq 0.74$ (dense) and occurs at concentration $c\simeq
2.27$\,mM, { which is consistent with the ``knee'' observed in the
  equilibrium surface tension (\ref{tens_eq})}. This is { also} close
to the concentration at which the dynamic surface tension curves
deviate more clearly from DLA behavior.~\ref{phi_0} shows the
theoretical equilibrium coverage as a function of concentration, { as
  calculated by numerically solving eq~\plainref{gamma_eq}} for the
same parameter values used in the fit of~\ref{tens_eq}. The range of
surface coverage and bulk concentrations corresponding to the
transition is rather small, { and, furthermore, sensitive to the
  fitted value of $\beta$. (See dashed lines in \ref{tens_eq}.)} This
can possibly explain why a transition has not been detected in the BAM
experiments.

\subsection{Kinetics at short times}

At short times ($t_{\rm age} < 5\,$s) but yet larger than $\taud$, and for low surfactant concentrations, as is demonstrated in~\ref{c124}, the dynamic surface tension curves fit quite well the asymptotic time dependence of a DLA process\cite{HD_jpc96,HD_csa01}.
\begin{equation}
  \Delta\gamma(t\gg\taud) \simeq
  \Delta\gamma_\eq\left(1-\sqrt{\frac{\taud}{t}} \right).
\label{diff}
\end{equation}
The fitted values for $\taud$ and $\Delta\gamma_\eq$ are given in~\ref{tab_taud}.

\begin{table}[tbh]
\begin{tabular}{|c|c|c|c|}
\hline
& & & \\
$c$ (mM)~ & ~$\taud$ (ms)~ & ~$\Delta\gamma_\eq$ (mN/m)& $D \times 10 ^6$ cm$^2$ s$^{-1}$ \\
\hline
$2.16$ & $1.05$ & $-5.56$ &$2.3 $\\
$3.24$ & $0.36$ & $-9.68$ &$3.1$\\
$4.32$ & $0.52$ & $-13.1$ &$1.2$\\
$3.8$ (MPT2) & $0.053$ & $-12.0$ &$10$\\
\hline
\end{tabular}
\caption[]{\textsf{Fitted values for the diffusion time $\taud$ and
the equilibrium reduction in surface tension $\Delta\gamma_\eq$
using eq~\plainref{diff}. We have also included the value obtained from~\ref{MPT2}.
Note that in this case the adsorption times were not corrected by the apparatus function $\lambda$.}} \label{tab_taud}
\end{table}

According to the theory\cite{HD_csa01} $\taud$, which characterizes the relaxation of $\Delta\gamma$ in a DLA process, is given by

\begin{equation}
  \taud = \frac{\phi_{0,\eq}^4}{\phi_\rmb^2}\frac{a^2}{\pi D}\quad ,
\label{taud}
\end{equation}
where $D$ is the bulk diffusion coefficient of the surfactant
molecule.  From the known value of the surfactant concentration, its
molar weight $308 \,{\rm g/l}$ and density $0.684\,{\rm g/l}$, we can
estimate the volume fraction, $\phi_{\rm b}$. At sufficiently high
concentration, after the diffusion step, the surface tension reaches
values corresponding to an almost saturated monolayer (constant slope
region of the experimental $\Delta\gamma_\eq$ as a function of $\log
c$ in \ref{tens_eq}).  Therefore, for $c>2$ mM we will assume that
$\phi_{0,\eq} \simeq 1$.  This leads to
\begin{equation}
  D \simeq \frac{1}{\phi_b^2} \frac{a^2}{\pi \tau_d}\quad.
\label{a4D}
\end{equation}
The value of the molecular size, $a\simeq 7.2$\,\AA\ was obtained in
the previous section from the fit to the equilibrium isotherm and is
about equal to the value estimated from the molecular volume, $a^3$,
where $a \simeq 8$\,\AA.
The calculated $D$ values are listed in \ref{tab_taud} for $c$ values
between 2 and 4\,mM. These $D$ values are a bit lower than the known
value for DTAB diffusivity in solution\cite{DTABdiff}, $D\simeq
6\times 10^{-6}$ cm$^2$/s. This may indicate that the diffusion into
the sub-surface region is slightly slower than the diffusion in the
bulk.  All the curves in~\ref{c124} clearly show that the adsorption
kinetics is not controlled by diffusion at times larger than $t >
1$\,s as will be discussed in the following section.

\subsection{Kinetics at longer times}

As is evident from~\ref{c124},
the dynamic surface tension curves at longer times deviate substantially from the DLA
behavior and cannot be well fitted with an inverse-square-root temporal decay. This becomes more pronounced as the concentration increases. In the insets of~\ref{c124} we re-plot the data on a semi-logarithmic scale, demonstrating that the final relaxation to equilibrium is exponential,

\begin{equation}
  \gamma(t) - \gamma_\eq \sim {\rm e}^{-t/\tauk}.
\label{kinetic}
\end{equation}
This relaxation is consistent with a kinetically limited adsorption (KLA), where the process is hindered by adsorption barriers.~\ref{tab_tauk} lists the fitted values of the relaxation time $\tauk$. These values are 2--3 orders of magnitude larger than the diffusion times $\taud$ listed in~\ref{tab_taud}.

\begin{table}[tbh]
\begin{tabular}{|c|c|}
\hline
~$c$ (mM)~ & ~$\tauk$ (s)~ \\
\hline
$1.08$ & $1.06$ \\
$2.16$ & $0.86$ \\
$3.24$ & $0.23$ \\
$4.32$ & $1.24$ \\
$6.48$ & $0.27$ \\
$8.64$ & $0.05$ \\
$10.8$ & $0.33$ \\
$13$ & $1.3$ \\
\hline
\end{tabular}
\caption[]{\textsf{Fitted values for the kinetic relaxation time
$\tauk$.}} \label{tab_tauk}
\end{table}

From the theory\cite{HD_jpc96,HD_csa01} we expect

\begin{equation}
 \tauk = \taud {\exp}\left(-\alpha-\beta\phi_{0,\eq} + 2 e\widehat{\psi}/k_\rmB T\right),
\label{tauk}
\end{equation}
where $\taud$ has been defined in eq~\plainref{taud}, and
$\widehat{\psi}=(\psi_0+\psi_a)/2$ is the average of the equilibrium
electrostatic potentials at the surface and sub-surface layers. From
eqs~\plainref{taud} and~\plainref{tauk} with $\phi_{0,\eq}\simeq 1$ we get
$e\widehat{\psi}/k_\rmB T = [\alpha+\beta+\ln(\pi a^4 D c^2 \tauk)]/2$.
Using the fitted values of $\tauk$ (\ref{tab_tauk}), $D = 6\times 10^{-6}$\,cm$^2$/s,
and the fitted equilibrium values $\alpha\simeq 11.8$, $\beta\simeq
7.6$ and $a\simeq 7.2$\AA, we obtain for the four higher bulk
concentrations, respectively, four rather similar values for the
average surface potential: $e\widehat{\psi}/k_\rmB
T\simeq 12.35,\; 11.73,\; 12.88$, and $14.96$.

On the other hand, the Poisson-Boltzmann theory\cite{HD_jpc96} yields
\begin{equation}
  {\rm e}^{e\psi_0/k_\rmB T}=\left[b\phi_{0,\eq}+\sqrt{(b\phi_{0,\eq})^2+1}\right]^2
  \simeq (2b\phi_{0,\eq})^2,
\end{equation}
where $b(a,\phi_\rmb)$ has been defined below eq~\plainref{gamma_eq}.
Taking $\phi_{0,\eq}\simeq 1$ and $a\simeq 7.2$ \AA, we get for the
four higher bulk concentrations, respectively,\cite{d8,d9}
$e{\psi_0}/k_\rmB T \simeq 8.4,\; 8.1,\; 7.9$, and $7$.
{ Since the potential at the subsurface layer must be smaller than the
surface one, $\psi_a<\psi_0$, we should expect to find
$\psi_0>\widehat{\psi}$.} The values found for $\psi_0$, based on the
equilibrium Poisson-Boltzmann (PB) theory, are comparable, but smaller
than the aforementioned potential barriers $\widehat{\psi}$ inferred
from the dynamic surface tension measurements. This may reflect some
inadequacies of the PB theory to account for all the experimental
results as reported here.

\section{Conclusions}

In the present work we have analyzed the equilibrium and dynamic surface tension of DTAB at the water-air interface for several concentrations, all below the critical micellar concentration (\emph{cmc}).

At short times DTAB adsorbs in a diffusion-limited process (DLA) with
a $\sim t^{-1/2}$ temporal relaxation and a diffusion coefficient $
D\simeq 10^{-6}$\,cm$^2$/s. At longer times, the DLA behavior is
followed by a kinetically limited adsorption (KLA) with an exponential
relaxation, as predicted by the theory of Ref~\citenum{HD_jpc96} for
salt-free surfactant solutions.  For the higher concentrations, the
dynamic surface tension exhibits an intermediate plateau, followed by
a final, exponential relaxation occurring over time scales of several
seconds. This behavior is similar to the one reported earlier for the
adsorption of SDS at a water-oil interface in the absence of
salt\cite{anne}.

The experiments indicate that the adsorption of DTAB at the water-air
interface undergoes a qualitative change in behavior in the
concentration range, $2\le c \le 3$\,mM.  We could not detect the
predicted phase transition using Brewster angle microscopy (BAM), but
this could arise either because the surface domains are smaller than
the optical resolution or because the transition occurs at
concentrations other than those investigated (the predicted
concentration range is very narrow, see~\ref{MPT2}).

The kinetically limited relaxation is related to an electrostatic
barrier created by the charged surface. The relevance of
electrostatics is supported by the observed strong effect of added
salt on the adsorption kinetics. Our analysis indicates, however, that
the Poisson-Boltzmann theory does not fully account for the observed
potential barriers. { The discrepancy might originate from effects
  related to the finite size of ions concentrated close to the surface
  \cite{DavidPRL}. We note that such effects were included in previous
  studies \cite{Battal2003,Valkovska2004}, where equilibrium
  measurements were well fitted by a van der Waals isotherm with a
  Stern layer of bound counterions \cite{Kralchevsky1999}. Similar
  modifications of the current theory may also affect predictions
  concerning the occurrence of a surface phase transition during the
  adsorption process.}

Another mechanism that might be invoked in principle involves the
possible formation of pre-micellar aggregates at concentrations closer
to (but below) the bulk solution {\em cmc}. In a previous theoretical
study\cite{Mohrbach}, it has been shown that the adsorption kinetics
in the presence of micelles exhibits an exponential relaxation, with a
relaxation time that is related to the exchange of surfactants between
aggregates and the surface. However, the exchange time for small chain
surfactants is of the order of a microsecond\cite{Lang,ZanaBook}, and
is too fast to be seen in our experiments. Furthermore, one expects
added salt to promote aggregate stability and, thus, to push the
system toward KLA, in disagreement with our measurements.

In conclusion, the current work brings further insight into possible
mechanisms of ionic surfactant adsorption at fluid interfaces. Further
theoretical and experimental work is necessary to fully understand
this important phenomenon that controls the dynamic behavior of the
interfaces.

\bigskip
{\em Acknowledgement:~~~} Support from Centre National de Recherche Spatiale (CNES) and European Space Agency (ESA)
under MAP projects AO-99-075 and 108, the Israel Science Foundation (ISF) under grant no. 231/08 and the US--Israel Binational Science
Foundation (BSF) under grant no. 2006/055 is gratefully acknowledged.  The experiments with the in-house made MBP instrument has been performed
in the Facultad de Ingenieria of Buenos Aires and we thank Prof. D. Kurlat for his support. The experiments with the commercial MBP instrument has been performed in the Universidad Complutense and we thank Prof. R. Rubio and Prof. F. Ortega for their support as well. HR thanks the MICIN for a RyC contract and DA thanks the Triangle de la Physique, France (POMICO project No.
2008-027T) for a travel grant.

\newpage


\end{document}